# Structural Reinforcement in Mechanically Interlocked Two-Dimensional Polymers by Suppressing Interlayer Sliding

Ye Yang[1,12], André Knapp[2,12], David Bodesheim[3,12], Alexander Croy[4], Mike Hambsch[5], Chandrasekhar Naisa[1,8], Darius Pohl[6], Bernd Rellinghaus[6], Changsheng Zhao[7], Stefan C. B. Mannsfeld[5], Gianaurelio Cuniberti[3,10]*, Zhiyong Wang[1,8]*, Renhao Dong[1,9]*, Andreas Fery[2,11]*, and Xinliang Feng[1,8]*

[1]Center for Advancing Electronics Dresden & Faculty of Chemistry and Food Chemistry, Technische Universität Dresden, Dresden, Germany

[2]Institute of Physical Chemistry and Polymer Physics, Leibniz Institute of Polymer Research Dresden, Dresden, Germany

[3]Institute for Materials Science and Max Bergmann Center for Biomaterials, TU Dresden, Dresden, Germany

[4]Institute of Physical Chemistry, Friedrich Schiller University Jena, Jena, Germany

[5]Center for Advancing Electronics Dresden & Faculty of Electrical and Computer Engineering, Technische Universität Dresden, Dresden, Germany

[6]Dresden Center for Nanoanalysis (DCN), Center for Advancing Electronics Dresden (cfaed), TUD Dresden University of Technology, Dresden, Germany

[7]College of Polymer Science and Engineering, State Key Laboratory of Polymer Materials Engineering, Sichuan University, Chengdu, China

[8]Max Planck Institute for Microstructure Physics, Halle (Saale), Germany

[9]Key Laboratory of Colloid and Interface Chemistry of the Ministry of Education, School of Chemistry and Chemical Engineering, Shandong University, Jinan, China

[10]Dresden Center for Computational Materials Science (DCMS), TU Dresden, Dresden, Germany

[11]Chair for Physical Chemistry of Polymeric Materials, Technical University Dresden, Dresden, Germany

[12]These authors contributed equally: Ye Yang, André Knapp, David Bodesheim

*E-mail: gianaurelio.cuniberti@tu-dresden.de; wang.zhiyong@tu-dresden.de; renhaodong@sdu.edu.cn; fery@ipfdd.de; xinliang.feng@tu-dresden.de

**Preserving the superior mechanical properties of monolayer two-dimensional (2D) materials when transitioning to bilayer and layer-stacked structures poses a great challenge, primarily arising from the weak van der Waals (vdW) forces that facilitate interlayer sliding and decoupling. Here, we discover that mechanically interlocked 2D polymers (2DPs) offer a means for structural reinforcement from monolayer to bilayer. Incorporating macrocyclic molecules with one and two cavities into 2DPs backbones enables the precision synthesis of mechanically interlocked monolayer (MI-M2DP) and bilayer (MI-B2DP). Intriguingly, we have observed an exceptionally high effective Young's modulus of 222.4±51.0 GPa for MI-B2DP, surpassing those of MI-M2DP (130.1±15.2 GPa), vdW-stacked MI-M2DPs (2×MI-M2DP, 8.1±0.9 GPa) and other reported multilayer 2DPs. Modeling studies demonstrate the extraordinary effectiveness of mechanically interlocked structures in minimizing interlayer sliding (~0.1 Å) and**

**energy penalty (320 kcal/mol) in MI-B2DP compared to 2×MI-M2DP (~1.2 Å, 550 kcal/mol), thereby suppressing mechanical relaxation and resulting in prominent structural reinforcement.**

Two-dimensional (2D) materials (e.g., graphene,[1] transition metal dichalcogenide,[2] 2D polymers (2DPs) and their layer-stacked 2D covalent organic frameworks (COFs)[3,4]), as well as their van der Waals (vdW) heterostructures have exhibited fascinating physical properties that confer a range of advantages for novel electronic devices, energy conversion, and storage.[5,6] Of particular interest are their mechanical properties, which not only lay the foundation for manufacturing, integration, and performance in potential applications,[7,8] but also allow for the manipulation of the electronic, magnetic, and optical properties of 2D materials through strain engineering.[9-11] However, as a general dilemma faced by most multilayer 2D materials, the inherently weak vdW interaction presents a critical dependence on sliding and decoupling between the layers, leading to substantial structural relaxation and loss of mechanical property as the number of layers increases.[12-14] For instance, effective Young's modulus ($E_{Young}$) of graphene exhibits a progressive reduction with increasing layers, declining from 890 GPa for the monolayer to 390 GPa for the bilayer, 50 GPa for the trilayer, and 30 GPa for the multilayer (> 4 layers), respectively.[15] This limitation curtails the effectiveness of modulating the fundamental properties of 2D materials, and their integration into flexible devices.[16-18] Recently, theoretical and experimental studies have unveiled a substantial enhancement in interlayer interaction and a reduction in the interlayer distance by incorporating $sp^3$ bonds in bilayer graphene through chemical functionalization and interlayer bonding of C-C atoms.[19,20] Nevertheless, the introduction of $sp^3$ bonds simultaneously induces geometric imperfections, leading to partial degradation of the mechanical properties of bilayer graphene, specifically in terms of tensile strength and $E_{Young}$. Hence, reinforcing the interlayer interactions of 2D materials while effectively preserving their exceptional in-plane mechanical properties is still a significant challenge.

In this study, we discover a structural reinforcement in the transition from monolayer to bilayer 2DPs, characterized by the incorporation of macrocyclic molecules (MCMs) in the backbones. We show that the cooperative assembly between host (i.e. cucurbit[8]uril (**CB8**) and nor-seco-cucurbit[10]uril (**ns-CB10**) MCMs) and guest (i.e. 1,1′-bis(4-aminophenyl)-[4,4′-bipyridine]-1,1′-diium chloride (**V-2NH₂**)) molecules enables the precision synthesis of mechanically interlocked monolayer (**MI-M2DP**) and bilayer (**MI-B2DP**) 2DP films on the water surface. Characterizations through diffraction and X-ray scattering techniques reveal a hexagonal porous structure with an in-plane parameter of $a = b = 44.5$ Å. It is notable that the 'strain-induced elastic buckling instability for mechanical measurements' (SIEBIMM) technique[21,22] unveils that **MI-B2DP** possesses a record-high $E_{Young}$ of 222.4±51.0 GPa, in contrast to those of **MI-M2DP** (130.1±15.2 GPa), vdW-stacked **MI-M2DP** (**2×MI-M2DP**, 8.1±0.9 GPa) and other reported multilayer 2DPs (typically less than 50 GPa).[23,24] Theoretical calculations support these findings, demonstrating that at a certain strain of $\varepsilon$=0.03, **MI-B2DP** exhibits significantly reduced interlayer sliding (~0.1 Å) compared to **2×MI-M2DP** (~1.2 Å), leading to a decreased energy loss (320 kcal/mol versus 550 kcal/mol) associated with mechanical relaxation. These findings shed light on the fundamental understanding of layer-dependent structural relaxation in 2D materials and provide potential avenues to address the challenges of preserving mechanical properties from monolayer to multilayer materials.

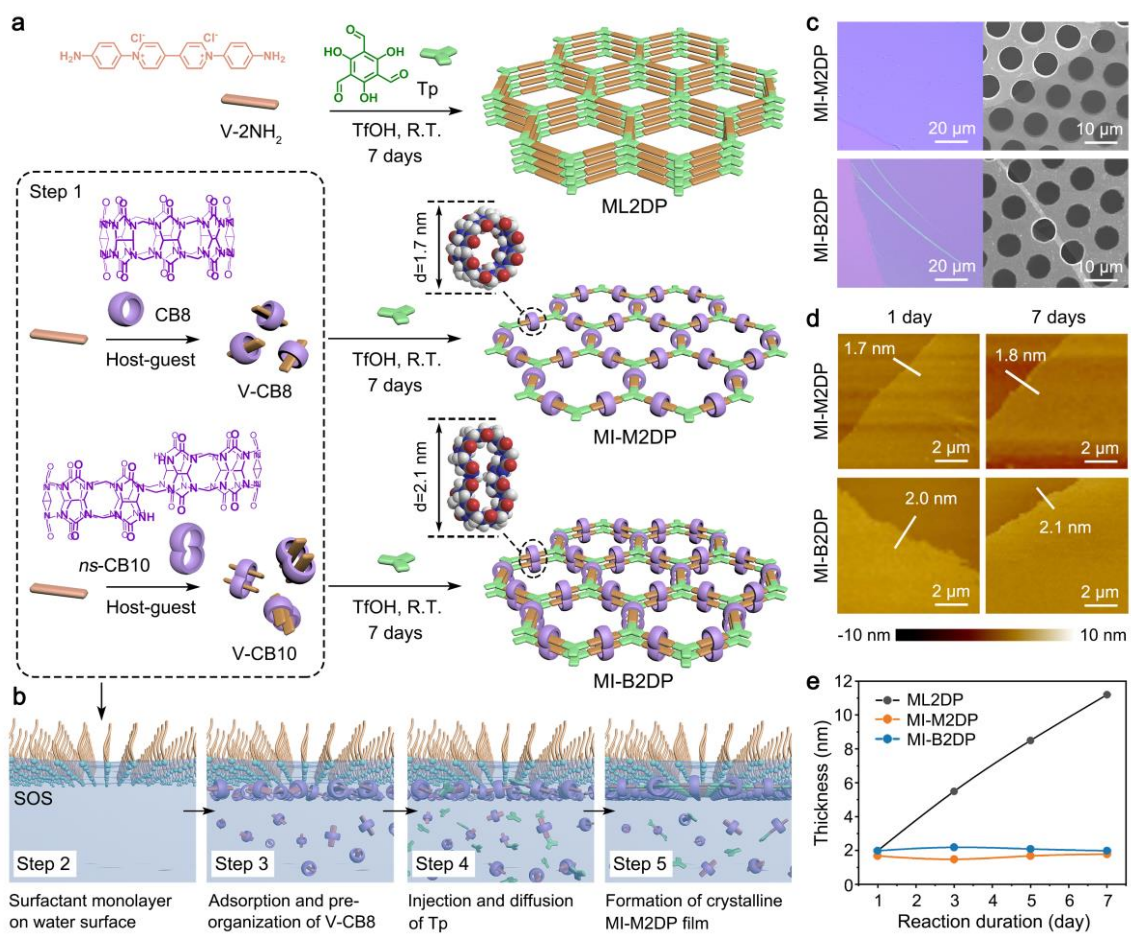

**Fig. 1| On-water surface synthesis and characterization of MI-M2DP and MI-B2DP. a,** Reaction schemes illustrating the synthesis of **ML2DP, MI-M2DP** and **MI-B2DP** via (A$_2$+B$_3$)-type 2D polycondensation, and host-guest assembly process between **V-2NH$_2$** and macrocyclic host (i.e. **CB8** (outer diameter, 1.7 nm), **ns-CB10** (outer diameter, 2.1 nm)) (Step 1). **b,** Schematic illustration of the synthetic procedure through the SMAIS method, involving Steps 2-5. **c,** OM image of **MI-M2DP** and **MI-B2DP** on SiO$_2$/Si substrates and SEM image of **MI-M2DP** and **MI-B2DP** on copper grids with a hole area of ~20 μm$^2$. **d,** AFM images of **MI-M2DP** and **MI-B2DP** on SiO$_2$/Si substrates with respect to reaction time. The thicknesses of the films along the white line are marked. **e,** Thickness of **ML2DP**, **MI-M2DP** and **MI-B2DP** on SiO$_2$/Si substrates with respect to reaction time.

## Results

**On-water surface synthesis of MI-M2DP and MI-B2DP.** MCMs[25,26] have garnered significant interest as supramolecular scaffolds for constructing linear polymers that include a mechanical bond and crosslinked polymer networks through host-guest chemistry.[27,28] Building on this concept, we propose that MCMs containing two or more host cavities for guest molecules could enable the synthesis of mechanically interlocked 2DPs with controlled layer numbers and enhanced interlayer interactions, thereby spatially suppressing their interlayer sliding and preventing the structural relaxation.[29] The synthesis of **MI-M2DP** and **MI-B2DP** using surfactant monolayer-assisted interfacial synthesis (SMAIS)[4,30] method on the water surface is illustrated in Figs. 1a,b. First, monomers **V-CB8** and **V-CB10** were synthesized in aqueous solutions by incorporating **CB8** and **ns-CB10**[31,32] (Supplementary

Figs. 1-4) into the backbone of **V-2NH₂**, respectively, as building blocks for creating **MI-M2DP** and **MI-B2DP** (Step 1). The successful formation of **V-CB8** and **V-CB10** was confirmed by UV-vis absorption and $^1$H NMR studies (Supplementary Figs. 5-7).[33] Then, a monolayer of sodium oleyl sulfate (SOS) was prepared on the water surface (Step 2), followed by the injection of 1 ml mixed aqueous solution of trifluoromethanesulfonic acid (TfOH, 7.4 µmol) and **V-CB8** (2.4 µmol, or **V-CB10** for **MI-B2DP** synthesis) into the water subphase (pH≈1.3). After the adsorption of monomers on the water surface for 2 h (Step 3, Supplementary Figs. 8-10), 1 ml aqueous solution of 2,4,6-trihydroxybenzene-1,3,5-tricarbaldehyde (**Tp**, 1.6 µmol) was added to the sublayer of the system to initiate the 2D polycondensation via Schiff-base reaction (Step 4). The polymerization was then kept undisturbed at room temperature for 1 day, affording a pale-yellow film with a lateral size of 28 cm$^2$ on the water surface (Step 5, Supplementary Fig. 11).

**Structural characterizations of MI-M2DP and MI-B2DP.** The attenuated total reflection-Fourier transform infrared (ATR-FTIR) spectroscopy shows that the stretching vibration of N-H (~3,323 cm$^{-1}$) from **V-CB8** and **V-CB10**, and -CHO (~1,640 cm$^{-1}$) from compound **Tp** completely disappeared after polycondensation (Supplementary Figs. 12 and 13), suggesting the complete conversion of monomers into 2DPs. Compared to multilayer 2DP without using MCMs (**ML2DP**), the characteristic FTIR peaks of -CH₂- (2,945 cm$^{-1}$) and C=O (1,716 cm$^{-1}$) from **V-CB8** and **V-CB10** monomers can be observed in **MI-M2DP** and **MI-B2DP**, supporting the successful embedding of **CB8** and **ns-CB10** into the 2DP networks. Furthermore, the chemical structure and composition of **MI-M2DP** and **MI-B2DP** were confirmed by surface-enhanced Raman and X-ray photoelectron spectroscopy (XPS) characterizations (Supplementary Figs. 14-18, Table S1). The energy dispersive X-ray (EDX) mapping also reveals a homogenous distribution of C, N, O, and F in both 2DP films (Supplementary Figs. 19 and 20).

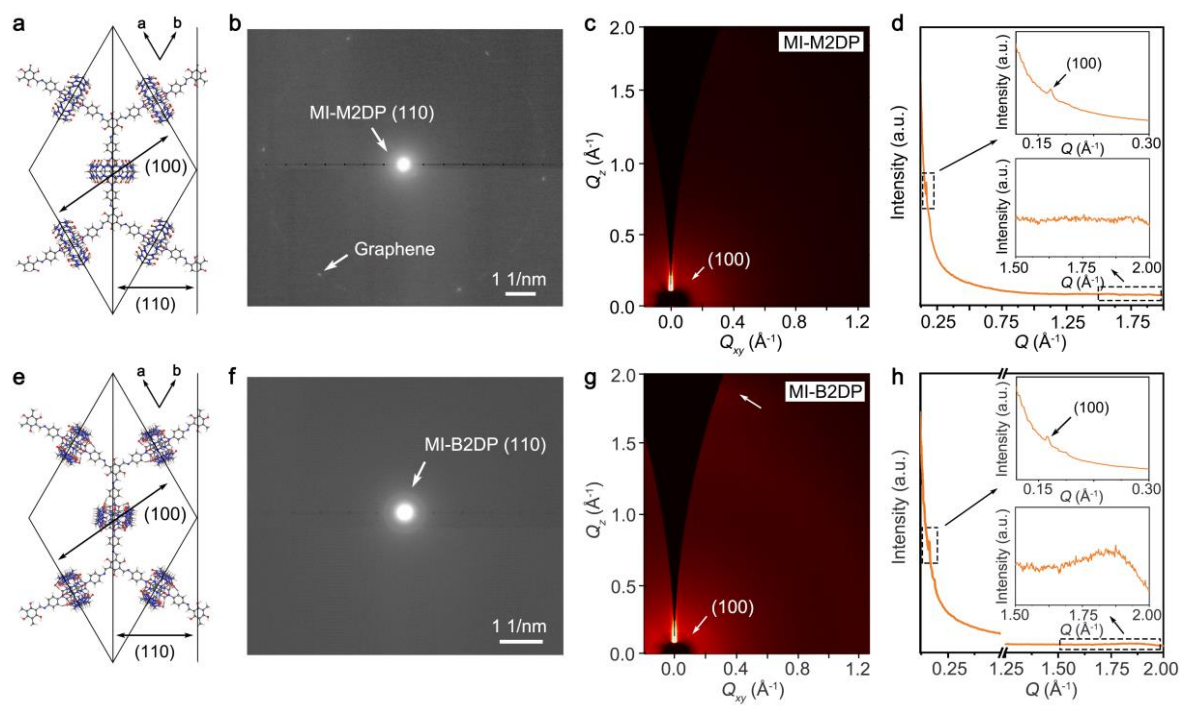

**Fig. 2| Structural characterizations of MI-M2DP and MI-B2DP films. a**, Schematic

illustration of **MI-M2DP** unit cell. The (100) and (110) planes of **MI-M2DP** are marked with black arrows. **b**, SAED pattern of **G/MI-M2DP/G** with a sandwich structure, which enables reducing the damage from the electron beam. **c,d**, GIWAXS pattern (**c**) and the integrated intensity profile (**d**) of **MI-M2DP** film. $Q$ represents the integrated scattering vector of in-plane ($Q_{xy}$) and out-of-plane ($Q_z$). **e**, Schematic illustration of **MI-B2DP** unit cell. The (100) and (110) planes of **MI-B2DP** are marked with black arrows. **f**, SAED pattern of **MI-B2DP** film. **g,h**, GIWAXS pattern (**g**) and the integrated intensity profile (**h**) of **MI-B2DP** film. $Q$ represents the integrated scattering vector of in-plane ($Q_{xy}$) and out-of-plane ($Q_z$).

The optical microscopy (OM) and scanning electron microscopy (SEM) images show the macroscopically homogeneous feature of **MI-M2DP** and **MI-B2DP** films (Supplementary Fig. 21). As shown in Figs. 1c, Supplementary Fig. 22 and 23, both **MI-M2DP** and **MI-B2DP** films can suspend over the holes (lateral size, ~20 µm$^2$) on a transmission electron microscopy (TEM) grid without rupturing, indicative of excellent mechanical stability.[34] Atomic force microscopy (AFM) analysis of **MI-M2DP** and **MI-B2DP** films show the root mean square (RMS) roughness of 0.18 nm and 0.27 nm in an area of 10×10 µm$^2$. The thicknesses of **MI-M2DP** and **MI-B2DP** are determined to be ~1.7 and 2.1 nm, respectively, aligning with the anticipated values for the monolayer and bilayer structures (Figs. 1d and Supplementary Fig. 24).[35] In contrast to the observed increase in thickness over time for **ML2DP** (from 2.0 nm for 1 day to 11.2 nm for 7 days), the thicknesses of **MI-M2DP** and **MI-B2DP** are maintained (Figs. 1e and Supplementary Fig. 25-29), suggesting that the bulky MCMs hinder the growth of layer-stacked structures on the water surface.

To investigate the crystal structure of **MI-M2DP** and **MI-B2DP**, we conducted TEM and synchrotron-based grazing-incidence wide-angle X-ray scattering (GIWAXS) measurements. The selected area electron diffraction (SAED) pattern obtained from a graphene/**MI-M2DP**/graphene (**G/MI-M2DP/G**) sandwich structure shows a weak diffraction ring at 0.45 nm$^{-1}$ (i.e. $d$-spacing of 22.2 Å) attributed to the (110) crystal plane (Figs. 2a,b and Supplementary Fig. 30-32),[36] indicating a polycrystalline nature. To probe the macroscopic structural order of **MI-M2DP**, we further performed GIWAXS measurement on a 20-layer **MI-M2DP** film prepared through the layer-by-layer (LBL) assembly. The in-plane reflection ring at $Q_{xy}$=0.17 Å$^{-1}$ (i.e. $d$-spacing of 37.0 Å) agrees well with the (100) plane of **MI-M2DP** (Fig. 2c,d), confirming its well-ordered in-plane molecular structure. Similar characterizations were then carried out for **MI-B2DP** samples (Fig. 2e). The SAED pattern shows an obvious reflection at 0.45 nm$^{-1}$ (i.e. $d$-spacing of 22.2 Å), which can be assigned to the (110) plane of **MI-B2DP** (Figs. 2f and Supplementary Fig. 33). The GIWAXS pattern displays a diffraction ring at $Q_{xy}$=0.17 Å$^{-1}$ (i.e. $d$-spacing of 37.0 Å), corresponding to the (100) in-plane parameter of **MI-B2DP**. Additionally, a weak reflection peak at $Q_z$=1.86 Å$^{-1}$ suggests an interlayer distance of 3.4 Å between **MI-B2DP** bilayers (Fig. 2g,h).[37] These results validate the successful synthesis of well-ordered **MI-M2DP** and **MI-B2DP** films, lending further credence to the feasibility of utilizing MCMs for precisely tuning the out-of-plane structure of 2DPs.

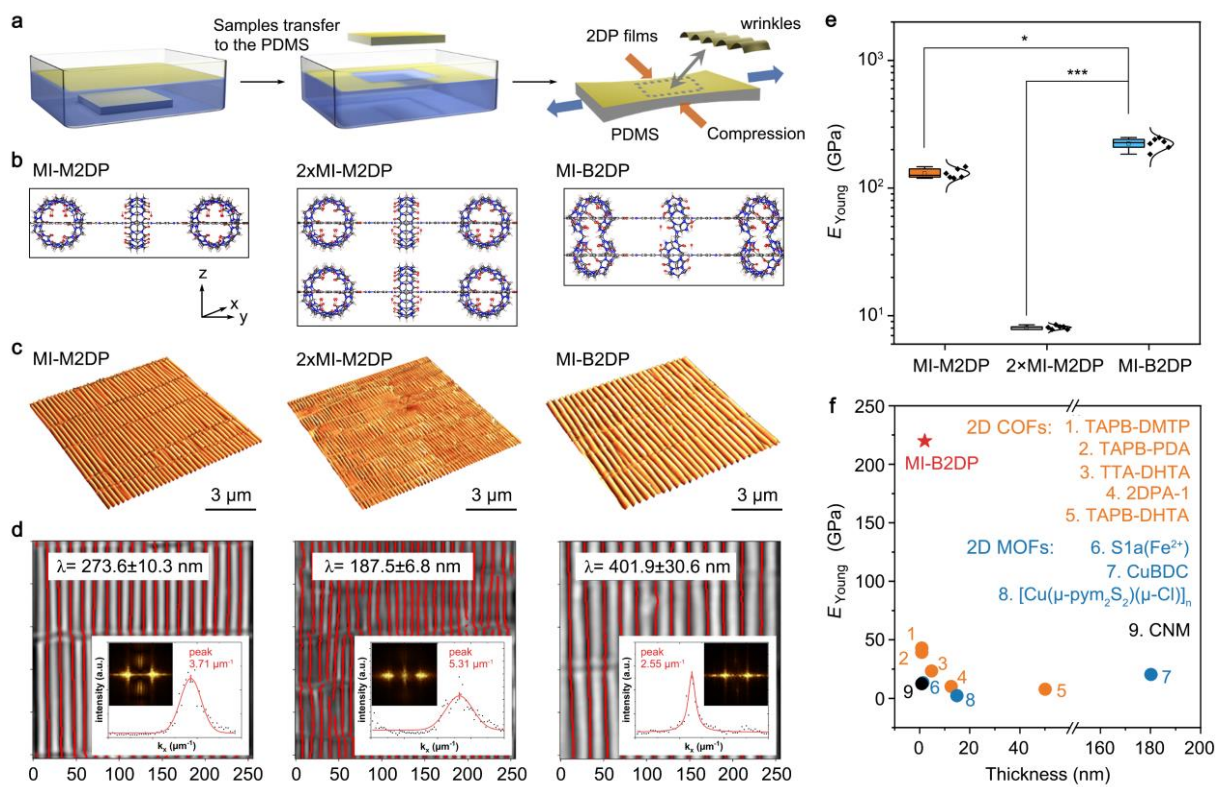

**Fig. 3| Mechanical properties of MI-M2DP, 2×MI-M2DP and MI-B2DP. a**, Schematic illustration of the transfer of 2DP films onto a plain PDMS and the lateral compression of PDMS during uniaxial transverse stretching, inducing monolayer wrinkling. **b**, Side view of **MI-M2DP**, **2×MI-M2DP** and **MI-B2DP**. **c**, AFM topographical images of large-scaled wrinkle pattern ($10 \times 10$ μm$^2$) for **MI-M2DP**, **2×MI-M2DP** and **MI-B2DP** at 10% compressive strain. **d**, Wavelength calculation images for **MI-M2DP**, **2×MI-M2DP**, and **MI-B2DP** based on the line-wise wavelength and amplitude calculation process. Integrated intensity profile after 2D Fourier-transformation along $k_x$ direction (insert). **e**, $E_{\text{Young}}$ of **MI-M2DP**, **2×MI-M2DP**, and **MI-B2DP**. Asterisks indicate significant differences (*P < 0.1, ***P < 0.001). All values are expressed as mean ± SD, n=6. **f**, Mechanical property comparison of **MI-B2DP** with the reported layer-stacked 2D COFs, 2D MOFs and carbon nanomembranes (CNM, ~2-450 layers).

**Mechanical properties of MI-M2DP and MI-B2DP.** To determine the mechanical properties of **MI-M2DP** and **MI-B2DP**, the SIEBIMM technique was employed.[21,22] The synthesized 2DP films were horizontally transferred onto a polydimethylsiloxane (PDMS) elastomeric support and strained using a motorized strain device, resulting in the formation of a regular wrinkling pattern perpendicular to the strain direction in 2DPs (Fig. 3a). Three samples, including **MI-M2DP**, two-layer stacked **MI-M2DP** (**2×MI-M2DP**) and **MI-B2DP**, were measured to investigate the impact of interlayer interactions on the mechanical properties (Fig. 3b). AFM topographical images show regular wrinkle patterns for all samples (Fig. 3c), indicating their high quality and suitability for SIEBIMM. The wavelengths of the wrinkles in **MI-M2DP**, **2×MI-M2DP** and **MI-B2DP**, calculated by a Python-based calculation method and cross-checked 2D Fourier-transformation,[38] were 273.6±10.3, 187.5±6.8 and 401.9±30.6 nm, respectively (Fig. 3d and Supplementary Fig. 34). The $E_{\text{Young}}$ was thus evaluated using the regular wrinkling wavelength, film thicknesses (Supplementary Fig. 35 and Table S2), and the

mechanical stiffness of the PDMS substrate (2.06 MPa), as described by the following equation.[39,40]

$$E_{\text{Young\_f}} = \frac{3E_{\text{Young\_s}}}{(1-v_s^2)} \left(\frac{\lambda}{2\pi h}\right)^3 (1-v_f^2) \qquad 1$$

where $\lambda$ represents the wrinkle wavelength, $h$ is the thickness of the upper film, $v$ is Poisson's ratio, $E_{\text{Young\_f}}$ is the $E_{\text{Young}}$ of film and $E_{\text{Young\_s}}$ is the $E_{\text{Young}}$ of substrate. The resulting $E_{\text{Young}}$ values for **MI-M2DP** ($E_{\text{MI-M2DP}}$=130.1±15.2 GPa) and **MI-B2DP** ($E_{\text{MI-B2DP}}$=222.4±51.0 GPa) are comparable to those of other well-known monolayer 2D materials (Fig. 3e and Supplementary Fig. 36), such as $MoS_2$ (270 GPa),[41] $WS_2$ (272 GPa),[42] $Ti_3C_2T_x$ (333 GPa),[43] and Graphene (1 TPa).[44] Furthermore, as a result of its unique mechanically interlocked structure, **MI-B2DP** film exhibits an exceptionally high $E_{\text{Young}}$ compared to the reported layer-stacked 2D COFs, 2D metal-organic frameworks (2D MOFs), and carbon nanomembranes (~2-450 layers, Fig. 3f).[23,24] These results suggest the superior mechanical properties of **MI-B2DP** films.

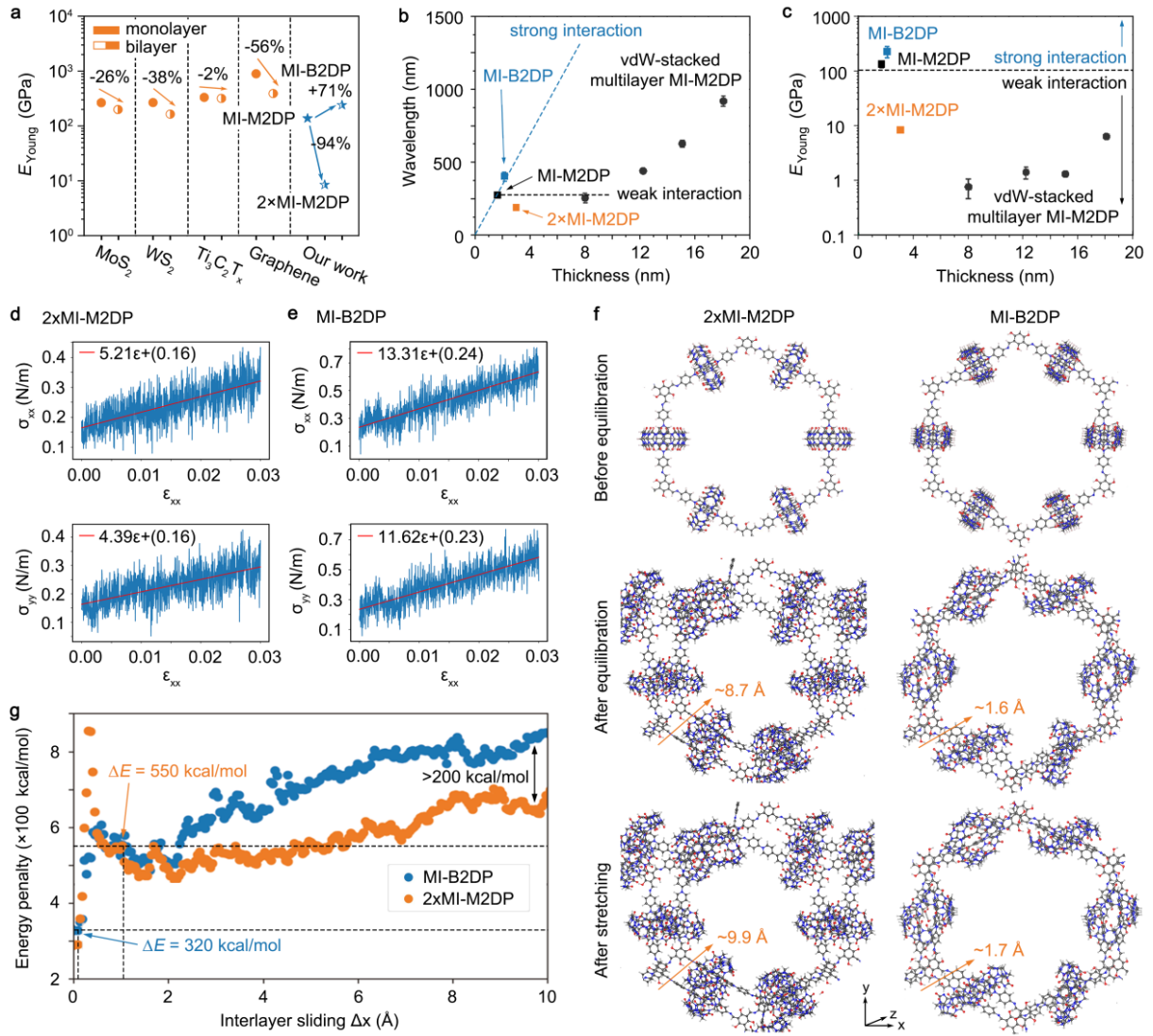

**Fig. 4| Interlayer behavior in vdW-stacked 2×MI-M2DP and MI-B2DP. a**, Mechanical property comparison of **MI-M2DP**, **MI-B2DP** and **2×MI-M2DP** with those of state-of-the-art monolayer and bilayer 2D materials. **b,c**, Thickness-dependent wavelength results λ(h) (**b**)

and $E_{\text{Young}}$ (**c**) for **MI-M2DP**, **MI-B2DP**, vdW-stacked **2×MI-M2DP**, and vdW-stacked multilayer **MI-M2DP** (~5-30 layers) at 10% compressive strain. **d,e,** Calculated in-plane stress-strain curves along xx and yy of **2×MI-M2DP** (**d**) and **MI-B2DP** (**e**). **f,** Simulated chemical structure of **2×MI-M2DP** (left) and **MI-B2DP** (right) during equilibration and stretching. **g,** Energy barrier of **MI-B2DP** and **2×MI-M2DP** upon interlayer sliding.

**Wavelength and theoretical studies of interlayer sliding in bilayer 2DPs.** It is notable that the $E_{\text{Young}}$ of the mechanically interlocked **MI-B2DP** increased by 71% compared to **MI-M2DP**, while the $E_{\text{Young}}$ of the vdW-stacked **2×MI-M2DP** decreased by 94% (to 8.1±0.9 GPa). This stands as a unique phenomenon since, typically, when stacking monolayer into bilayer 2D materials, interlayer sliding impedes the collective engagement of the layers in the stress-strain process, leading to mechanical relaxation and subsequent degradation of their $E_{\text{Young}}$ (Fig. 4a).[15,41-43] However, the mechanically interlocked structure tightly integrates the layers, forming a composite layer that enables simultaneous contributions to the mechanical strength of 2DP films. To evaluate the interlayer behavior during mechanical property measurement, we further performed images morphology analysis. The high-resolution AFM images display well-ordered wrinkle patterns in **MI-M2DP** and **MI-B2DP**, indicating strong interlayer adhesion. In contrast, a locally irregular pattern was observed in **2×MI-M2DP**, revealing interlayer sliding and decoupling during compression (Supplementary Fig. 37). Furthermore, the obtained wavelength versus thickness dependency λ(h) also enables to act as a qualitative indicator to characterize the interlayer adhesion behavior of multilayer samples. The wavelengths of **MI-M2DP** and **MI-B2DP** follow a linear and orthogonal λ(h) relation (blue line, Fig. 4b) as the converted wrinkling equation.[45]

$$\lambda = 2\pi h \left( \frac{E_{\text{Young}\_f}/\left(1-v_f^2\right)}{3\,E_{\text{Young}\_s}/\left(1-v_s^2\right)} \right)^{1/3} \qquad\qquad 2$$

This phenomenon provides evidence for the robust interlayer coupling in **MI-B2DP**, which effectively constrains both the in-plane and out-of-plane motion of the bilayer, thereby influencing the wrinkling process. In contrast, the vdW-stacked **2×MI-M2DP**, with a thickness of ~3.0 nm, exhibits an opposing behavior characterized by a significant reduction in wavelength due to weak interlayer adhesion. During compression, the bottom layer of the **2×MI-M2DP**, which is well adhered to the PDMS substrate, initiates wrinkling, resulting in a wavelength comparable to that of the **MI-M2DP** film. However, the vdW-stacked top layer of **2×MI-M2DP** does not contribute to the wrinkling process. Instead, it undergoes mechanical relaxation due to interlayer sliding, merely following the given wrinkled surface without delamination in the best case. The same behavior was observed for the vdW-stacked multilayer **MI-M2DP** (~5-30 layers) films as their thicknesses increase from ~3.0 to 18.1 nm, leading to the pronounced differences in calculated $E_{\text{Young}}$ (Fig. 4c). These results demonstrate the significant impact of interlayer interactions on the mechanical properties of 2D materials.

To gain deeper insights into the structural reinforcement in **MI-B2DP** enabled by MCMs, we further performed classical molecular dynamics (MD) simulations to investigate the stress-strain behavior of **2×MI-M2DP** and **MI-B2DP**. As shown in Figs. 4d,e, and Supplementary Fig. 38, **MI-B2DP** possesses a significantly enhanced in-plane stress response to applied strain, characterized by a higher 2D $E_{\text{Young}}$ ($E_{\text{Young}\_2D}$ = 3.2 N/m) compared to **2×MI-M2DP** ($E_{\text{Young}\_2D}$ = 1.5 N/m), which supports the experimental findings (Fig. 3e). It is notable that for **2×MI-M2DP** stacked by weak vdW forces, a significant interlayer sliding (~8.7 Å) was already observed even during the initial equilibration, whereas the layers of **MI-B2DP** interlocked by MCMs remained nearly AA stacked (~1.6 Å) (Fig. 4f, Videos S1-S4). Furthermore, after stretching ($\varepsilon$=0.03), additional interlayer sliding (~1.2 Å) occurred in **2×MI-M2DP** compared to **MI-B2DP** (~0.1 Å). These results demonstrate that the weakly vdW-stacked **2×MI-M2DP** is more prone to sliding during stress loading. We further calculated the energy penalty for interlayer sliding in these two bilayer systems to elucidate the resulting energy loss from equilibration to the stretched state (Fig. 4g). Apart from an initial transient spike, up to a sliding distance of ~2 Å, the resulting energy penalties for **2×MI-M2DP** and **MI-B2DP** are comparable. For larger distances (>2 Å), the sliding energy of **MI-B2DP** is consistently higher (by up to 200 kcal/mol) compared to **2×MI-M2DP**, indicating its stronger interlayer coupling. In addition, at the strain of $\varepsilon$=0.03, the larger sliding distance (~1.2 Å) in **2×MI-M2DP** results in a substantially higher energy penalty (550 kcal/mol) than that of **MI-B2DP** (~0.1 Å, 320 kcal/mol), thereby giving rise to a greater modulus loss. These results reveal that the interlocked structure confers strongly coupled layers to **MI-B2DP**, ultimately endowing it with superior mechanical properties than **MI-M2DP** and **2×MI-M2DP**.

**Discussion**

In summary, we have demonstrated a structural reinforcement in the mechanically-interlocked 2DPs through the incorporation of MCMs in the backbones. The embedded MCMs with one and two cavities guide the precision synthesis of monolayer (**MI-M2DP**) and bilayer (**MI-B2DP**) films on the water surface. We have shown that the mechanically interlocked structure in **MI-B2DP** suppresses the interlayer sliding and tightly integrates the two layers as a composite layer. Such features enable both layers to contribute simultaneously to the mechanical strength of 2DP films, leading to a record-high $E_{Young}$ of **MI-B2DP** (222.4±51.0 GPa), compared to those of **M-2DMIP** (130.1±15.2 GPa), vdW-stacked **2×MI-M2DP** (8.1±0.9 GPa), and other reported layer-stacked 2D COFs, 2D MOFs and carbon nanomembranes (<50 GPa). Modeling analysis further demonstrates that compared to **2×MI-M2DP** (~1.2 Å), the interlayer sliding of **MI-B2DP** (~0.1 Å) is significantly restricted at the strain of $\varepsilon$=0.03, resulting in the reduced energy penalty (320 kcal/mol versus 550 kcal/mol) associated with the mechanical relaxation. Therefore, our results violate the common negative correlation between layer number and $E_{Young}$ observed in 2D materials, underscoring the significance of MCMs in reinforcing the structure of 2DPs. Our findings not only pave the way for reinforcing mechanical properties of 2D materials from monolayer to multilayer, but also offer insights into the exploration of novel mechanically interlocked microstructure physics.

**Methods**

**Synthesis of MI-M2DP film.** 1 mg of **V-2NH₂**, 3.24 mg of **CB8** and 1 ml of TfOH (7.4 µmol) aqueous solution were added into a glass bottle, and sonicated for 30 min to prepare the **V-CB8**. Then, 50 ml of Milli-Q water was added to a crystallization dish (60 ml, diameter ~ 6 cm). 10 µl of SOS solution (1 mg/ml in chloroform) was spread on the water surface by a micron injector. The chloroform on the water surface evaporated in 30 min. Subsequently, 1 ml of the prepared **V-CB8** solution was gently injected into the water subphase by using a syringe. After 2 h, 1 ml aqueous solution of **Tp** (1.6 µmol) was injected into the system for the polymerization. The reaction was kept undisturbed at room temperature for 1 days. The synthesized **MI-M2DP** was transferred on different substrates by the horizontal dipping method. Then, it was washed with chloroform, ethanol, and Milli-Q water, and dried at 50 ºC for 1 h.

**Synthesis of MI-B2DP film.** 1 mg of **V-2NH₂**, 1.99 mg of ***ns*-CB10** and 1 ml of TfOH (7.4 µmol) aqueous solution were added into a glass bottle, and sonicated for 30 min to prepare the **V-CB10**. Then, 50 ml of Milli-Q water was added to a crystallization dish (60 ml, diameter ~ 6 cm). 10 µl of SOS solution (1 mg/ml in chloroform) was spread on the water surface by a micron injector. The chloroform on the water surface evaporated in 30 min. Subsequently, 1 ml of the prepared **V-CB10** solution was gently injected into the water subphase by using a syringe. After 2 h, 1 ml aqueous solution of **Tp** (1.6 µmol) was injected into the system for the polymerization. The reaction was kept undisturbed at room temperature for 1 days. The synthesized **MI-B2DP** was transferred on different substrates by the horizontal dipping method. Then, it was washed with chloroform, ethanol, and Milli-Q water, and dried at 50 ºC for 1 h.

**X-ray diffraction refinements**. The GIWAXS measurements were performed at BL11-NCD-SWEET at ALBA Synchrotron in Barcelona, Spain. The beam energy was 12.4 keV and the beam had dimensions of 30 µm vertically and 150 µm horizontally. The Rayonix LX255-HS area detector was positioned 300.3 mm from the sample, which was verified by measuring a chromium oxide reference. The incidence angle was set to 0.12° and four images with 1-2 s exposure were recorded. The GIWAXS data was corrected and analyzed using WxDiff.

**SIEBIMM technique**. The PDMS system Sylgard 184 (Dow Corning Ltd, Midland, USA) was mixed in a 10:1 mass ratio of oligomeric base to curing agent and degassed for 2 min at 2200 rpm with a Thinky ARE250 tumbling mixer (Thinky Corporation, Tokyo, Japan), respectively. The mixture was cast into a poly(styrene) case with a thickness of 2 mm and cured first at room temperature for 48 h followed by a post-curing step at 80 °C for 4 h. The cured PDMS was cut into 45×10×2 mm specimens. To activate the surface, a pretreatment step with 10 wt% aqueous HCl solution for 16 h was necessary to provide sufficient adhesion to the 2DPs.[46] These samples were used to lift out different synthesized 2DPs from the water surface to create the necessary bilayer system for wrinkling. To induce the wrinkling, the samples were uniaxially strained to 20% with a motorized strain device, which introduce a perpendicular compressive strain by the Poisson ratio of 10%. The resulting wrinkle pattern was analyzed with an AFM (FastScan, Bruker Corporation, Billerica, USA), which was operated in Tapping mode with NanoScope 9.3 software. A Tap300 cantilever (tip radius 15 nm, spring constant 40 N m⁻¹, nominal resonance frequency 300 kHz) was used for topographical measurements. The images were taken in sizes of 50×50 mm² and 10×10 mm² by using a resolution of 512×512 pixels. The wrinkle wavelength was determined using a Python-based (Python 3.0) wrinkle calculation script based on the topographical images.[38]

**Simulations of elastic properties.** The elastic properties were simulated using classical molecular dynamics simulations with the Large-scale Atomic/Molecular Massively Parallel Simulator (LAMMPS)[47] with a parametrization of the force-field Reax-FF[48] at 10 K. Stress-strain curves were obtained by straining the system along xx or yy respectively and recording the stresses. The interlayer shearing simulations were performed by shifting the two layers against each other and recording the total energy of the system. After an initial geometry relaxation, the system was equilibrated using the Nose-Hoover barostat without external pressure with a temperature damping of 30 fs and stress damping of 300 fs in a NpT ensemble. The equilibration was performed for 200000 timesteps. After the NpT equilibration, the system was equilibrated in a NVT ensemble for another 100000 timesteps. The equilibrated system was stretched along the x-direction at a strain rate of 0.000001/fs for 100000 timesteps. The stresses were recorded during the simulation and the strain components were calculated based on the relative change of the box length and width. Through the resulting stress-strain curves, the elements of the stiffness tensor were calculated. Based on the elements of a stiffness tensor, the 2D bulk, shear and $E_{\text{Young}}$ of a system of hexagonal symmetry were evaluated.[49] For the simulation of the interlayer shearing, first a NVT equilibration was performed for 100000 timesteps. Then, several carbon atoms at the top and at the bottom layer were fixed along x-direction and another short NVT equilibration of 10000 timesteps was performed. The fixed atoms of the two layers were subsequently shifted against each other along x with a velocity of 7.46 $10^{-4}$ Å/fs and -7.46 $10^{-4}$ Å/fs respectively over 100000 timesteps. During the interlayer shearing, the total energy of the system was recorded.

**Data availability**

The data that support the findings of this study are available from the corresponding authors on request.

18943-18951 (2022).

**Acknowledgements**
This work was financially supported by the EU Graphene Flagship (GrapheneCore3, no. 881603), an ERC starting grant (FC2DMOF, grant no. 852909), an ERC Consolidator Grant (T2DCP), a DFG project (2D polyanilines, no. 426572620), the European Union's Horizon 2020 research and Innovation programme under grant agreement N° 101008701, EMERGE project, Coordination Networks: Building Blocks for Functional Systems (SPP 1928, COORNET), H2020-MSCA-ITN (ULTIMATE, no. 813036), H2020-FETOPEN (PROGENY, 899205), CRC 1415 (Chemistry of Synthetic Two-Dimensional Materials, no. 417590517), SPP 2244 (2DMP), as well as the German Science Council and Center of Advancing Electronics Dresden. R.D. thanks Taishan Scholars Program of Shandong Province (tsqn201909047) and National Natural Science Foundation of China (22272092). Y.Y. gratefully acknowledges funding from the China Scholarship Council. The authors acknowledge the Dresden Center for Nanoanalysis at TUD. The authors thank Dr. P. Formanek for the use of the SEM facility and Dr. Y. Liu for the structural calculation of **ns-CB10**. The GIWAXS experiments were performed at the BL11-NCD-SWEET beamline at ALBA Synchrotron with the collaboration of ALBA staff. The authors would like to thank Dr. M. Malfois for the help in setting up the experiment.

**Author Contributions**
Z.W., R.D. and X.F. conceived and designed the project. Y.Y., Z.W. and C.N. contributed to the synthesis. Y.Y. performed OM, SEM, AFM, NMR, ATR-FTIR and UV-vis measurements. C.Z. performed XPS measurement. D.P., Z.W. and Y.Y. performed HRTEM imaging and SAED measurements, and the corresponding analysis. Y.Y., R.D., M.H. and S.M. performed the GIWAXS measurement and the analysis. Y.L. contributed to the theory calculations. A.K. and A.F. performed the mechanical property measurements and the analysis. D.B., A.C. and G.C. contributed to the mechanical property simulations. Y.Y., Z.W., R.D., A.K. and X.F. co-wrote the manuscript with contributions from all the authors.

**Competing interests**
The authors declare no competing interests.

**Additional information**
Supplementary information is available for this paper at https://
Correspondence and requests for materials should be addressed to G.C., Z.W., R.D., A.F. and X.F.
Reprints and permissions information is available at http://